\documentclass[]{spie}  
\pdfoutput=1
 
\usepackage{amsmath,amsfonts,amssymb}
\usepackage{graphicx}
\usepackage{cite}
\usepackage[colorlinks=true, allcolors=blue]{hyperref}
\usepackage{xcolor}

\title{Deep learning solutions to telescope pointing and guiding}

\author[a,c]{Jackson Zariski}
\author[a]{Kaitlin M. Kratter}
\author[b]{Sarah E. Logsdon}
\author[a]{Chad Bender}
\author[b]{Dan Li}
\author[b]{Heidi Schweiker}
\author[b]{Jayadev Rajagopal}
\author[b]{Bill McBride}
\author[b]{Emily Hunting}
\affil[a]{Steward Observatory, University of Arizona. (United States)}
\affil[b]{NSF National Optical-Infrared Astronomy Research Lab. (United States)}
\affil[c]{Department of Applied Mathematics, University of Arizona (United States)}

\authorinfo{Further author information: (Send correspondence to Jackson A. Zariski)\\: E-mail: jzariski@arizona.edu, Telephone: 1 425 295 4790}
 
\begin{document} 
\maketitle

\begin{abstract}
The WIYN 3.5m Telescope at Kitt Peak National Observatory hosts a suite of optical and near-infrared instruments, including an extreme precision, optical spectrograph, NEID, built for exoplanet radial velocity studies. In order to achieve sub ms$^{-1}$ precision, NEID has strict requirements on survey efficiency, stellar image positioning, and guiding performance, which have exceeded the native capabilities of the telescope’s original pointing and tracking system. In order to improve the operational efficiency of the telescope we have developed a novel telescope pointing system, built on both a recurrent neural network and gradient-boosted tree, that does not rely on the usual pointing models (TPoint or other quasi-physical bases). We discuss the development of this system, how the intrinsic properties of the pointing problem inform our network design, and show preliminary results from our best models. We also discuss plans for the generalization of this framework, so that it can be applied at other sites.
\end{abstract}


\section{INTRODUCTION}
The WIYN telescope\footnote{The WIYN Observatory is a joint facility of the NSF's National Optical-Infrared Astronomy Research Laboratory, Indiana University, the University of Wisconsin-Madison, Pennsylvania State University, Purdue University and Princeton University.}, located at Kitt Peak National Observatory in Arizona, hosts a primary mirror measuring $3.5$m in diameter, with a total weight of approximately 35 tons. It currently hosts several facility instruments, including NEID, a high-precision optical fiber-fed spectrograph built to conduct an exoplanet radial velocity survey \cite{NEID_paper,doppler}. To meet its survey goals, NEID requires a high efficiency observing strategy in order to observe an average of approximately $15$ targets per night of operation. At present, survey efficiency is hampered by an imprecise, multi-stage target acquisition process. Targets are first acquired through a telescope mounted Star Tracker Camera, then centered on the NEID science fiber through multiple iterations with a secondary guider camera. This process not only takes time and operator input, but occasionally fails entirely when a target cannot be located in the wide-field Star Tracker camera.

\begin{figure}
    \centering
    \includegraphics[scale=0.5]{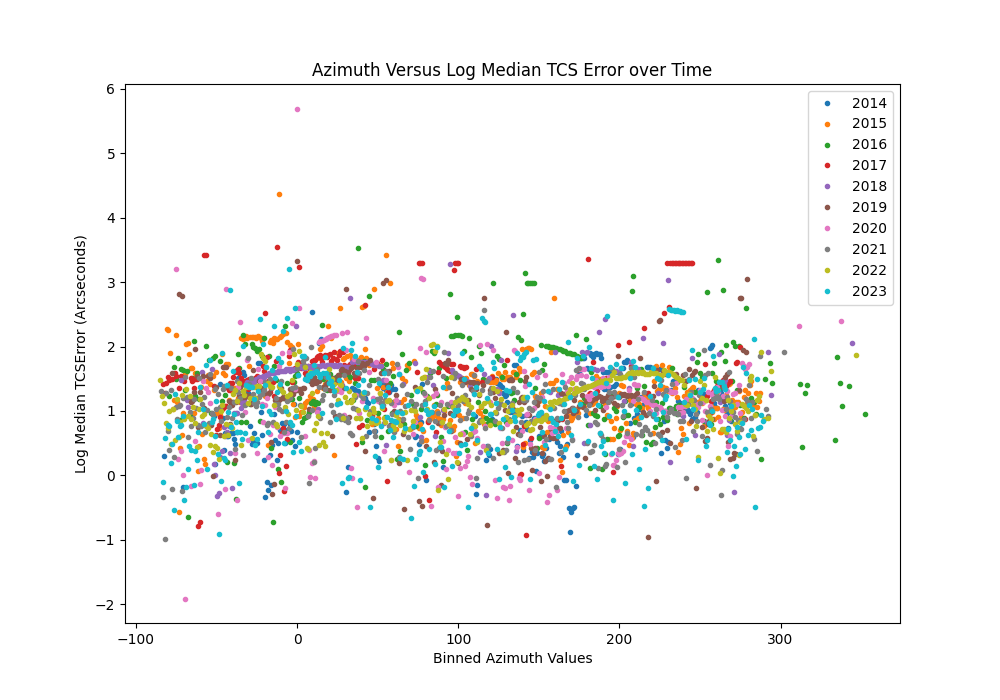}
    \caption{A historical look at how acquisition errors in right-ascension at the WIYN depend on mount azimuth. Each point represents the log of the error, measured in degrees, between the expected target right-ascension ($\alpha$), and that measured by the acquisition camera based on an astrometric database. Colors (see legend) indicate years in which the data is recorded. We can see noticeable, systematic error at certain coordinates, which we aim to correct with machine-learning.}
    \label{fig:azoffsets}
\end{figure}

The WIYN, like many professional telescopes, currently employs proprietary software, TPoint \cite{tpoint}, to generate a full-sky pointing solution. This regression model requires re-calibration multiple times per year, with telescope operators spending up to half a night acquiring benchmark stars across the sky. Re-calibration is required because pointing solutions drift due to myriad effects, such as motor wear, instrument re-mounting, and thermal expansion and contraction\cite{problempaper}. Moreover, the pointing model has highly non-uniform, time-dependent errors. In Fig. \ref{fig:azoffsets}, we show the historical typical pointing error as a function of telescope azimuth. Consequently on any given night, targets that happen to fall at high-error azimuths will likely be difficult to acquire. In its current implementation, these errors are neither systematically tracked nor reported to operators and engineers to alert users of a possible challenge (or mechanical issues).

To improve the WIYN's overall operating efficiency and in particular enable NEID to meet its design targets, we are developing a novel method for generating an adaptive, accurate, pointing model using various neural network architectures. In addition, though part of the broader category of machine-learning methods that are not necessarily `deep', we also employ the use of gradient-boosted trees for the best results. We have built a prototype model based on historical telemetry that can accurately predict $90\%$ of telescope acquisitions within roughly a quarter of an arcminute, compared to a typical error closer to $90$ arcseconds with the current system. Moreover, such a machine-learning informed model does not require that any additional observing time be dedicated to data collection or re-calibration, as requisite data is gained through normal operation, and model re-calibration can be done offline during daylight hours on a daily basis.

In Sec. \ref{sec:dlback} we provide an accessible summary of the machine-learning models we employ. In  Sec. \ref{sec:acq} we describe the development of our prototype pointing model using nearly a decade of historical telemetry. We then continue in Sec. \ref{sec:track} to describe ongoing efforts to generate a predictive model for target tracking during an observing sequence to reduce unnecessary telescope slews due to inaccuracies in the underlying pointing solution. We conclude in Sec. \ref{sec:summary} with a discussion of future directions for our work.
\label{sec:intro}  

\section{Background to Deep Learning}
\label{sec:dlback}
To introduce some of the primary machine-learning concepts that we employ, consider just the acquisition problem in the following, simplified context. Suppose we have a celestial target, written in celestial coordinates as $(\alpha, \delta)$, in addition to a set of values regarding certain weather/atmospheric conditions $W$ and a set of values referencing telescope-specific variables (guiding camera settings, time of observation, etc.). We represent these values as the set $P$. Our ultimate goal then is to numerically generate some function $f(\alpha, \delta, W, P)\to H$ where $H$ is the mathematical space of horizon coordinates bounded by the minimum and maximum azimuth and elevation values that can be obtained by the telescope. The function $f$ should seek to minimize the difference between the pixel-center of the acquisition system and the acquired image of the celestial object. To find said function numerically, a variety of learning techniques exist, which can generally be separated into general iterative regression techniques, and those requiring deep learning.

\subsection{Shallow Versus Deep Learning}
When discussing common machine-learning techniques, more specifically deep learning, we must first make the distinction between more modern neural networks and traditional statistical methods. In our context, the term `shallow regression' will refer to models with just two layers--an input layer and an output layer, though this nomenclature is not standard. Regression in this sense can take a variety of forms, but it generally entails developing a function as described above in the context of some linear combination of weighted features. Using the previous notation, we would generate the function $f$ as 
\begin{equation}
\label{eq:lcombo}
f(\textbf{x}) = a(\ell_i x_i), ~~~ \textbf{x}\in \cup\{\alpha, \delta, W, P\}.
\end{equation}
Here $a(\cdot)$ is some function (called an activation), often a linear combination in the case of most linear regression schemes, though polynomial regression is also common for situations requiring more complexity. Deriving the coefficients $\ell_i$ that generate a function for the best possible fit of the data is a non-trivial task with a variety of possible avenues for `training'. As an example, the current system at the WIYN and many other telescopes (TPoint) uses a proprietary fitting method to determine various coefficients for a pre-defined function \cite{tpoint}. In addition, the TPoint solution is revised through a few `test' acquisitions following re-calibration--a process that requires substantial time on sky -- roughly half a night, multiple times per year.

To contextualize the deep learning process, we will first describe briefly the process for general iterative methods, which are trained and tested separately as follows. First available data is split into a training set and a testing set. Following this, a loss function $L$ is chosen to quantify the error in the model; mean-squared loss is a common choice. A set of weights $\{\ell^0_i\}$ are initialized and a loss is calculated following a pass of the training data; then, through some differentiation scheme (auto-diff, finite differences, etc.), the values for $\frac{\partial L}{\partial \ell_i}$ are derived. Finally, the initialized weights are updated through the operation \begin{equation}
\label{eq:update}
\ell_i^{k+1} = \ell_i^{k} - \eta \frac{\partial L}{\partial \ell_i^{k}},
\end{equation}
where $k$ is the current iterative step (epoch) and $\eta$ is some learning rate, which can be either constant or adjusted during training. Following a certain number of iterations, training ceases and the function $f(\textbf{x})$ is evaluated on the testing set with the current weights to measure performance.
\begin{figure}
    \centering
    \includegraphics[scale=0.5]{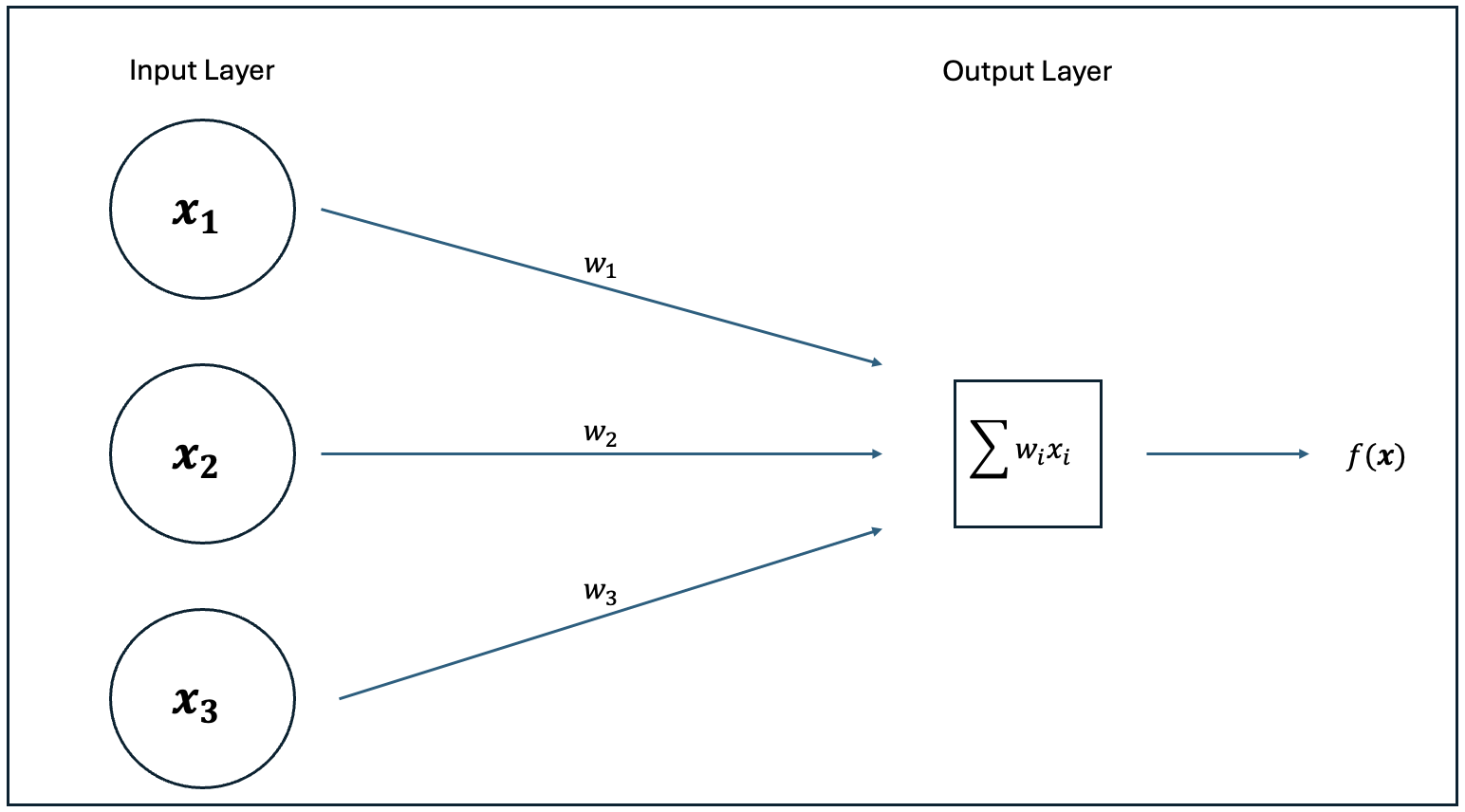}
    \caption{An example of a shallow network. The network accepts an input array $\textbf{x}$ comprised of three features, and generates a predictive function $f(\textbf{x})$. The activation used in this network is a simple linear combination. The values $w_i$ are the weights of the network.}
    \label{fig:shallownetwork}
\end{figure}
What differentiates the above learning scheme from the deep learning techniques that we currently employ is the inclusion of one or multiple `hidden layers' in the model. Now, instead of inputs to the network being weighted, activated and then directly returned as the output, they are instead passed to other layers, where further activations occur and the process continues. Gradient calculations for the loss function with respect to the weights are found through a procedure called back-propagation\cite{nn-book}. The partials for the weights connecting the output layer are computed first; then, through a clever implementation of auto-differentiation and dynamic programming, gradient calculation is propagated outwards towards the weights directly connecting the input layer to the first hidden layer. To apply this using the notation from above, we can represent the state of the $nth$ singular neuron in a deep network through some function $g_n$ where 
\begin{equation}
\label{eq:composition}
g_n(\textbf{x})=a_n(\cdots L_2a_1(L_1a_0(L_0\textbf{x}))), ~~~ \textbf{x}\in \cup\{\alpha, \delta, W, P\}.
\end{equation}
Here, each $L_i$ is the set of weights connecting to the corresponding previous neuron. So, $L_0$ is the set of weights directly connecting the input sequence to the first hidden layer, $L_1$ connects to the second, and so forth. Thus, deep learning clearly offers substantially more complexity than more traditional shallow methods, though with this comes an increase in time and memory requirements for both the training program and functional evaluation at runtime. The process described above is known as a simple feed-forward, densely connected network. While useful for certain tasks, our problem, which we can represent as a time-series forecast, benefits from a more complex architecture. 

\subsection{A Deeper Dive into Deep Learning: Recurrent Neural Networks}
While feed-forward neural networks are perfectly adept at modeling nonlinear relationships, when it comes to sequential data, the recurrent neural network stands out as a clear improvement. In essence, data is fed into a recurrent network grouped into sequences (such as sentences in a natural language processing model, or in our case, as sequential target acquisitions). Here, we classify an observation as a set of positional/environmental data points that is recorded at a single time-step. The first observation in the input sequence is weighted just like before; however, the output is used as a factor for the next observation in the sequence, in conjunction with the same weights. This continues throughout the sequence, and down through the network. Hence, a recurrent neural network acts as a `folded-up' feed-forward network that when unfurled, has both temporal weights and weights connecting neurons. While these extra temporal weights allow the model to learn connections between observations in a sequence (these networks can even be bidirectional where they learn from both directions), some problems can arise. For sequences of large length, the gradient is being multiplied by some repeated weights a large number of times. If these weights are small, $\ell<1$, the gradient will tend to vanish and approach zero, meaning that reaching a minimum for the loss function will take exponentially longer, or will be impossible. Meanwhile, with $\ell>1$, one encounters the exploding gradient problem, where the back-propagation process consistently over-shoots the minimum, again making convergence extremely difficult\cite{nn-book}. These problems can also arise in feed-forward networks (especially very deep ones), though due to the temporally linked nature of recurrent networks the problems are much more prevalent. However, some modifications to the recurrent unit exist to help rectify this issue. 
\begin{figure}
    \centering
    \includegraphics[scale=0.5]{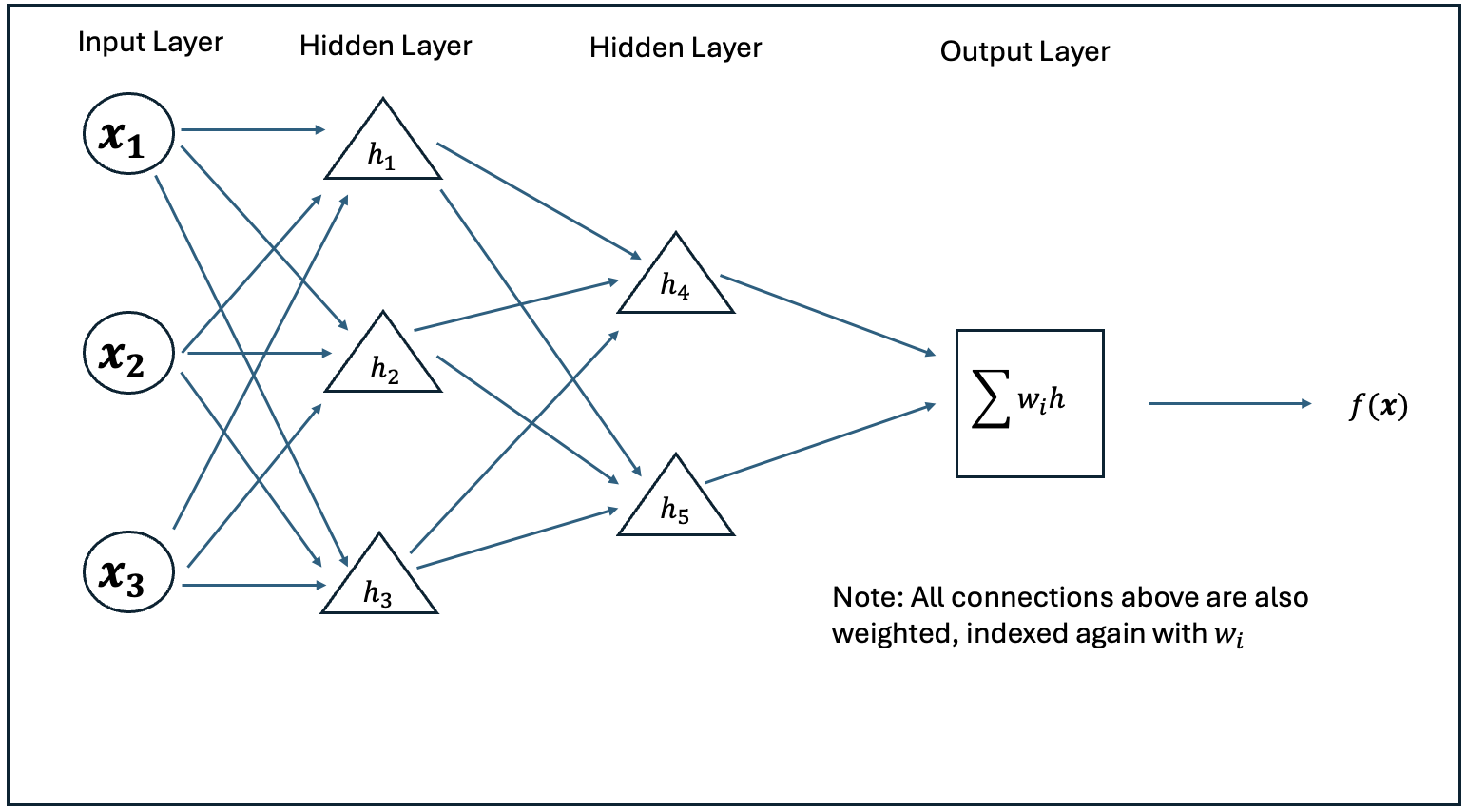}
    \caption{An example of a deep network with two additional hidden layers. The network again accepts an an input array $\textbf{x}$ comprised of three features, and generates a predictive function $f(\textbf{x})$. This time however, the input is propagated through nodes in the hidden layer, each with their own activations that add extra non-linearity to the model. Note that in general the choice of activation stays the same amongst nodes in the same layer.}
    \label{fig:deepnetwork}
\end{figure}

The problem with recurrent units described above is a problem of good `short-term' memory and poor `long-term' memory. Shorter sequences are exposed to less weight multiplications, and thus are less prone to vanishing or exploding. Longer sequences however are more exposed to this problem. Hence, we need a way to reliably train recurrent neural networks for sequences of any reasonable length. This is where we see the benefit of the Long Short-Term Memory (LSTM) unit, which has some extra variables in place to better mitigate this issue. Essentially, the LSTM cell contains a few running values (that are updated as sequences are propagated forward through the unit) which decide how much information to pass along without relying on compounding multiplications. In addition to the LSTM unit, a more recent adaptation is the Gated Recurrent Unit (GRU). With fewer operations, the GRU is less computationally expensive than the LSTM unit\cite{gru}, though still can be very effective. In general, during the training of a recurrent neural network, we treat the recurrent unit as a hyperparameter to examine, and try training the model using both types.

\subsection{Tuning Hyperparameters}
The word `hyperparameter' is very broad--it's used to reference anything from the number of hidden layers in a network to the type of units that compose the layers themselves. While there is no exact science to tuning hyperparameters (though automatic tuning is possible with a variety of different methods) one can also glean some intuition by analyzing how each parameter affects the network. Some commonly tuned for our purposes hyperparameters include:
\begin{itemize}
\item \textbf{Dropout (Regular and Recurrent)}:
Dropout is one of the primary ways to reduce over-fitting in a neural network. Regular dropout is included as a `layer' in the network, and assigned with some probability $p$. During the forward step of training, neurons in the hidden layer following the dropout layer are `dropped' with probability $1-p$. This process thins out the layer, and helps reduce the network from fitting too closely to the training data while making the model more general. Recurrent dropout works in a similar fashion as the method above, and is applied to the recurrent connections in a recurrent cell rather than the nodes themselves.
\item \textbf{Batch-Size}: This hyperparameter determines how we feed our training data through the network. If our batch-size is $b$, the network takes $b$ samples from the training set, runs it through the network, and then performs back-propagation to complete a part of training. This is then repeated for the next $b$ samples, and so on until all of the data is used. Note that higher batch sizes thus entail less training, and make the process less computationally expensive. However, they tend to add to the over-fitting problem, as they attract local minima solutions to the loss function, rather than global ones \cite{batchsize}. 
\item \textbf{Activation Function}:
This is the function applied to output values from a layer in the network. These functions are used for a variety of reasons, with one important reason being the mechanism to allow the network to output the correct class of data. In our networks, our final goal is to predict some continuous variable representing a desired targeting input for the telescope. As such, the values determined by our network must be continuous. In our case regarding time-series regression, the activation function in the output layer is simply $a(x)=x$, a linear activation. In addition to their use in the output layer, activation functions in the hidden layers of the network can add non-linearity to the model. In our case specifically, we utilize the Hyperbolic Tangent function (a popular choice for recurrent networks) and the Swish function, $\mathrm{swish}(x)=\frac{x}{1+e^{-\beta x}}, ~\beta>0$, to achieve this goal. These hidden layer activation functions are tuned as hyperparameters, since the function in the output layer is predetermined based on the purpose of the model.
\end{itemize}
\subsection{Gradient-Boosted Trees}
In addition to the neural network frameworks described above, we also implement a different type of machine-learning, known as gradient-boosted trees, specifically to aid in the acquisition problem. While technically not deep learning, gradient-boosted trees are still a powerful tool for regression tasks, especially in our case where we seek to be accurate on the order of thousandths of a degree. \cite{xgboost} The process of training these models also shares quite a few similarities to the optimization process undertaken by neural networks. Again, we choose a loss function and seek to optimize various parameters based on the residuals minimizing said loss function. Some of these parameters include how the decision tree splits input data into categories, the size of the weights on final outputs applied in a forest of decision trees, as well as the outputs of the trees themselves. Similar to back propagation, these factors are tuned by examining the residuals left from evaluating the training data and using them to find the gradient of said loss function with respect to the above parameters.

\section{Acquisition}
\label{sec:acq}

The WIYN presents a very well-suited environment for the development of a deep learning pointing solution, as we have both positional and environmental data logs dating back to 2014, as well as an already established targeting framework in TPoint to use as a baseline for improvement. The development of this system is built off of the Star Tracker camera, an Allied Vision Technologies GT1920 with a Tele-Xenar 2.2/70 lens fitted with an IR filter and a focal plane of 1936 X 1456 pixels. The field of view of the camera is approximately $7^\circ \times 5^\circ$, with a pixel-size of $13''$. The camera’s effective bandpass is 400-700 nm \cite{vimba}. Mounted at the top-end of the telescope, on a frame supporting the secondary mirror, the camera system uses a World Coordinate System to determine pointing by matching up stars in its field to known objects in astrometric catalogs. However, this `real' pointing location is only derived after the fact, when the telescope mount has already moved into position. Hence, we use the calculated celestial coordinate center from the star track center (in $\alpha, \delta$) as the ground truth positions that we aim to predict with our neural network training. These coordinates are hereafter referred to as the XY coordinates. With the Star Tracker camera as a base, our system will represent the pointing center for the telescope as the coordinate center of the image reported by the camera. Note that ultimately we desire a system that is trained to predict the mount encoder measured horizon coordinates, taking celestial coordinates as input. However, the historical dataset does not include the requisite data to analyze the model performance and errors in this direction, so to validate our method we choose to predict the celestial coordinates.  

Our training data consists of roughly $50,000$ target acquisitions over the course of eight years of observations. In addition to the initial horizon coordinates from the mount controller, we use features such as the angles of various port rotators on the telescope, the parallactic angle of the object we're observing, and the current pixel center on the camera. In addition, to bolster our model's ability to detect underlying dependencies of the pointing solution on weather, we include information on the dewpoint, temperature, wind-speed, and other climate-related features. Our target outputs are trained on the astrometrically determined celestial coordinates of the pixel center of the star tracking camera.

Though the acquisition data is not a uniform time series (targets are not acquired/reacquired at evenly spaced intervals), over the period of a night we do not expect too much deviation in the engineering and temperature variables we hope to capture in our network. In addition, some of these variables are included as features in our network as well, namely those that are quickly measurable relating to atmospheric conditions. 

\begin{figure}
    \centering
    \includegraphics[scale=0.4]{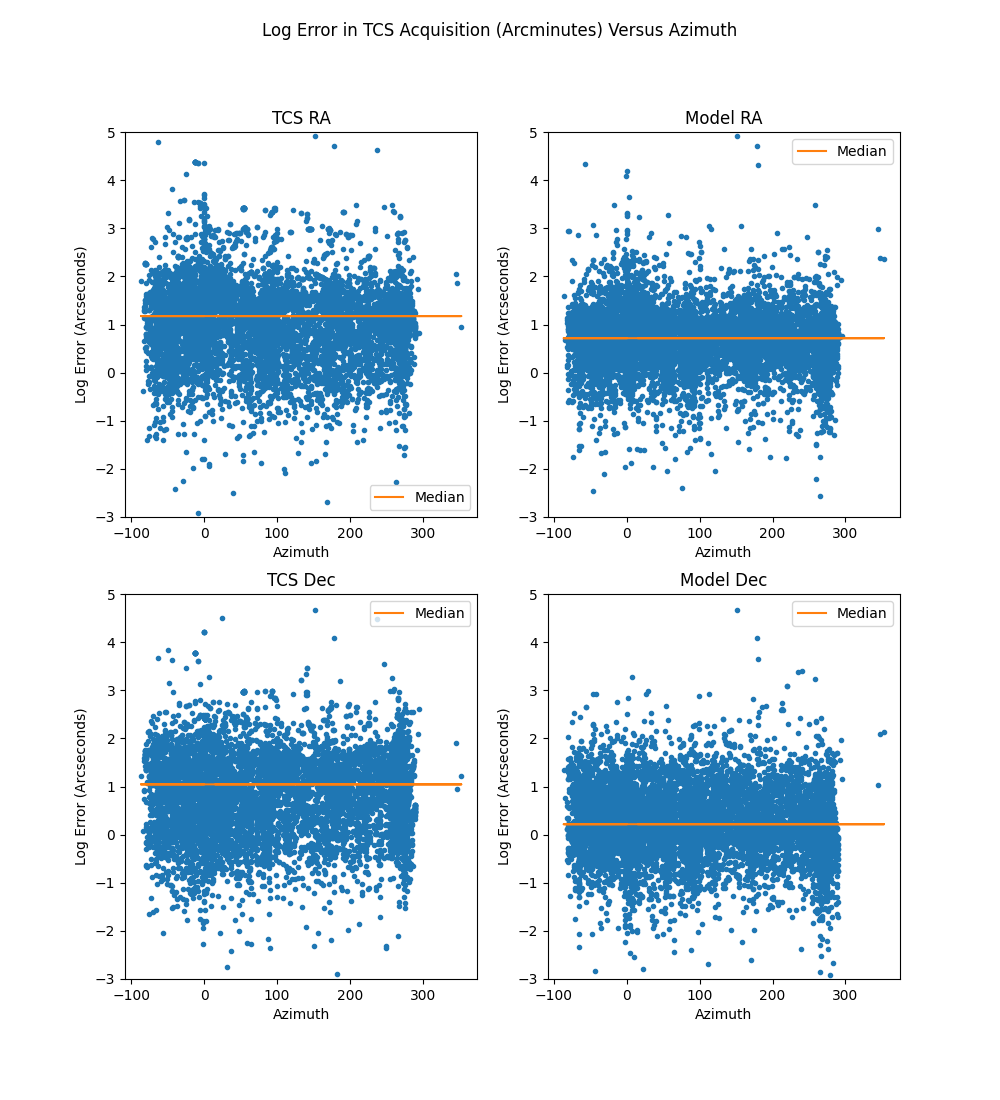}
    \caption{A comparison of how acquisition error of both the current WIYN system (TCS) and our model depends on azimuth. Other than some clear outliers, the deep learning model is less dependent on azimuth, and instead has more uniform accuracy throughout the entire coordinate interval. The orange line in this plot represents the median of the errors.}
    \label{fig:orangelineplot}
\end{figure}

\begin{figure}
    \centering
    \includegraphics[scale=0.65]{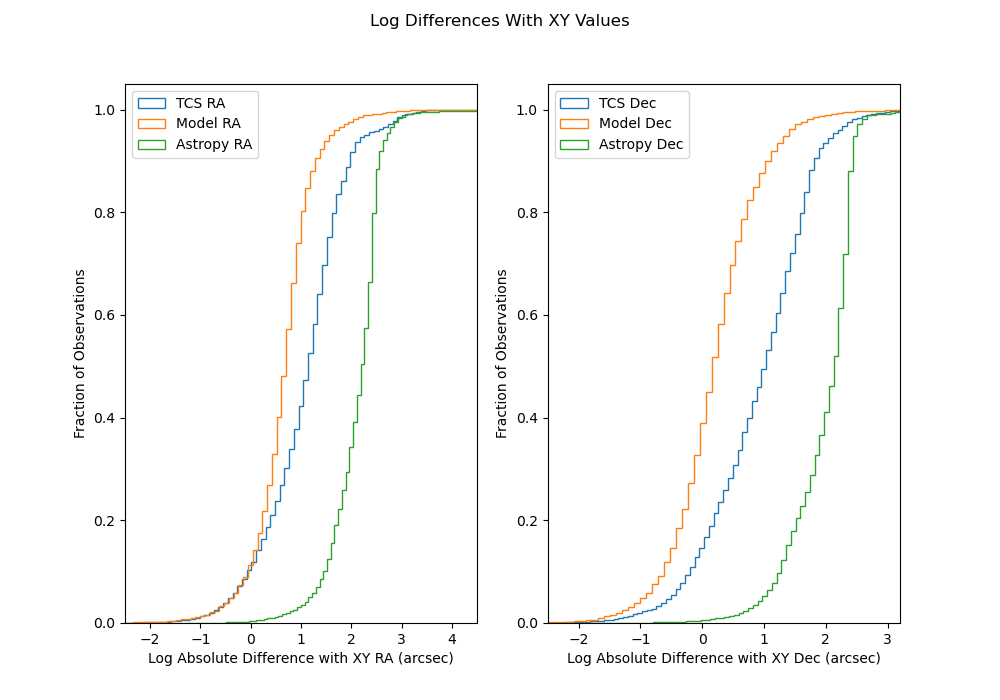}
    \caption{The cumulative distribution function representing the fraction of observations on the test set that fall below a given absolute error in each coordinate, shown with a logarithmic scale.   Here, the astropy measurements (green) come directly from the coordinate transform done via the astropy software\cite{astropy:2013}. This accounts for phenomena such as precession, nutation, etc. The TCS measurement (blue) comes from the current system at the WIYN and Model (orange) is our model built off of gradient-boosted trees. Our network displays a steep rise at lower absolute errors compared to the current system and astropy coordinates, indicating a substantial improvement in accuracy for most targets. Thus we conclude that machine learning can be an effective tool for target acquisition.}
    \label{fig:acqnetcelestial}
\end{figure}

\begin{table}[ht]
\caption{Median Absolute Differences with XY Celestial Coordinates during Target Acquisition} 
\label{tab:astratable}
\begin{center}       
\begin{tabular}{|l|l|l|l|} 
\hline
\rule[-1ex]{0pt}{3.5ex} Coordinate & TCS System & Recurrent Neural Network (NN) & XGBoosted Model\\
\hline
\rule[-1ex]{0pt}{3.5ex}  XY RA (Arcseconds) & 16.2 & 125.1 & 5.20 \\
\hline
\rule[-1ex]{0pt}{3.5ex}  XY Dec (Arcseconds) & 13.2 & 97.4 & 1.65 \\
\hline 
\end{tabular}
\end{center}
\end{table}
\newpage
As noted above, because of the nature of the current data available to us, we use the mount encoder horizon coordinates to predict celestial coordinates, in order to analyze the performance of our model relative to the existing telescope pointing solution. Summary metrics for our model's performance are shown in Table \ref{tab:astratable}. For programming these neural networks, we utilized the TensorFlow package in python which allows for the construction of networks with back-propagation done internally on the back-end\cite{tensorflow2015-whitepaper}. For the gradient-boosted tree model, we used the XGBoost package also found in Python\cite{xgboost}. We have validated that our model functions in reverse (i.e. using celestial coordinates to predict mount encoder horizon coordinates), but cannot generate a robust comparison with the existing pointing model in this direction. We illustrate the current performance of our model in Fig.~\ref{fig:acqnetcelestial}. Here, Model is the shorthand nomenclature for best model (XGBoost), TCS refers to the current pointing system\cite{tpm-paper}, and the `astropy' coordinates are shown for reference as the predictions from the python Astropy coordinate transformation package.\cite{astropy:2013,astropy:2018}  In addition, Fig.\ref{fig:orangelineplot} shows how our machine-learning framework is less prone to systematic errors in the azimuthal direction. Instead we see more uniform errors, and overall improved accuracy in comparison to the existing pointing solution. While the neural network as of now is not as accurate as our XGBoost model, it is still encouraging that we are able to achieve error on just the order of a few arcminutes using nothing but previously obtained Star Tracker data.

Even more encouraging, our best model has an error level that lies in the maximal precision afforded by the resolution of the star tracker camera ($\sim 6-7 ''$), and the historical data logs are missing some relevant data that might lead to an even more improved solution. For example we do not have data on which instrument was being used for each target, nor which instruments were mounted at the time, which affect the weight and balance of the telescope. Finally, while the Star Tracker camera is aligned to the telescope azimuth and elevation, it is not directly aligned to every instrument’s focal plane (some of which have field derotators), which could introduce additional errors. These features could all be included in future telemetry recorded with an eye towards developing new pointing models, and then easily ingested into the neural network and XGBoost training.

\section{Tracking--Synthetically-Generated Data}
\label{sec:track}
\subsection{Tracking Corrections with a Recurrent Neural Network}

In order to maintain the target on the fiber, NEID requires a pointing precision during observations of $0.05''$ for stars V = 12 mag or brighter under median seeing and wind conditions ($0.2''$ for 12 mag $< V<16$ mag)\cite{portadapter,rvebudget}. To maintain this accuracy, NEID employs two feedback loops while on target. The inner, high cadence loop ($27 Hz$) measures target offsets in the guider-camera, and adjusts a tip-tilt mirror at the image pupil for re-centering. The outer loop, with a cadence of $0.5 Hz$ sends offsets measured on the guider camera to the telescope control system, where pixel offsets, are converted to celestial coordinates, and subsequently horizon coordinates using the existing pointing solution\cite{fastguiding}. We highlight two potential sources of error, which we aim to address with predictive machine-learning models. First, both the inner and outer loop adjustments are inherently corrective, rather than predictive. By the time the telescope slew commands have been processed, the offset will be different from what has been computed. In addition, the multiple coordinate conversions (pixel to celestial to horizon) add errors into the model due to underlying inaccuracies in the pointing solution. We instead aim to train a predictive model that directly outputs the required adjustments in horizon coordinates, at the correct future timestamp when the telescope mount will execute the command, including latency in model compute time and system control. Crucially, as we discuss below, we design a self-correcting, or auto-regressive network, which is not only trained on historical data, but is informed and corrected by measuring pointing errors during tracking of each individual target.

Because implementation of this system requires specialized telemetry collection during scientific observations, we begin by validating this technique using a synthetically generated dataset.
Using the acquisition points recorded in the Star Tracker log as a base set of astronomical targets, we generate synthetic tracking data for each target with observing lengths ranging from a few minutes to half an hour\cite{NEID_paper}, with an exposure cadence of 1 Hz. We use astropy \cite{astropy:2013,astropy:2018} to covert the target ($\alpha, \delta$) from the Star Tracker log into a sequence of horizon coordinates as a function of the putative observing time. We then modulate this data to introduce both systematic and random errors to mimic deviations due to e.g. weather, telescope sag, load balance, and motor wear.

Our synthetic data takes the following form. First, define the random variable $X\sim\mathrm{Uniform}(u_1,u_2)$ (where $u_1$ and $u_2$ can be adjusted) with $n$-length vector $\textbf{X}$ sampled from this distribution, and a cumulative sum where 
\begin{equation}
\label{eq:sum}
\textbf{S}=\left(S_0,\cdots,S_n\right)~~~~S_i=\sum_{k=0}^i\textbf{X}_k.
\end{equation}
Next, let $\textbf{F}$ be a tensor of synthetic atmospheric features, where initial values are taken from a normal distribution and then propagated over a tracking period with some added noise to more closely mimic a stable yet slightly unpredictable pattern. Define the function $T:\textbf{F}\to\mathbb{R}$, which we generally write as a polynomial combination of synthetic features for smoothness purposes, though this can be changed to investigate other problems. In addition, we also include some systematic offsets at various intervals of horizon coordinate values to mimic machine and motor error. Calling this function $M$, then, the altered coordinates in our path take the form: 
\begin{equation}
\label{eq:totaleq}
(\textbf{Az}, \textbf{El}) = (\textbf{S}_1, \textbf{S}_2) + \epsilon_1 T(\textbf{F})+\epsilon_2M(\textbf{Az}, \textbf{El}),
\end{equation}
where $\epsilon_1,\epsilon_2\in\mathbb{R}$ depend on the magnitude of the features for the problem at hand and the desired amount of noise. These pairs of altered coordinates act as the output we train to predict. Henceforth, we will refer to the sequence of observations making up a tracking path in our data as $P$.

\begin{figure}
    \centering
    \includegraphics[scale=0.6]{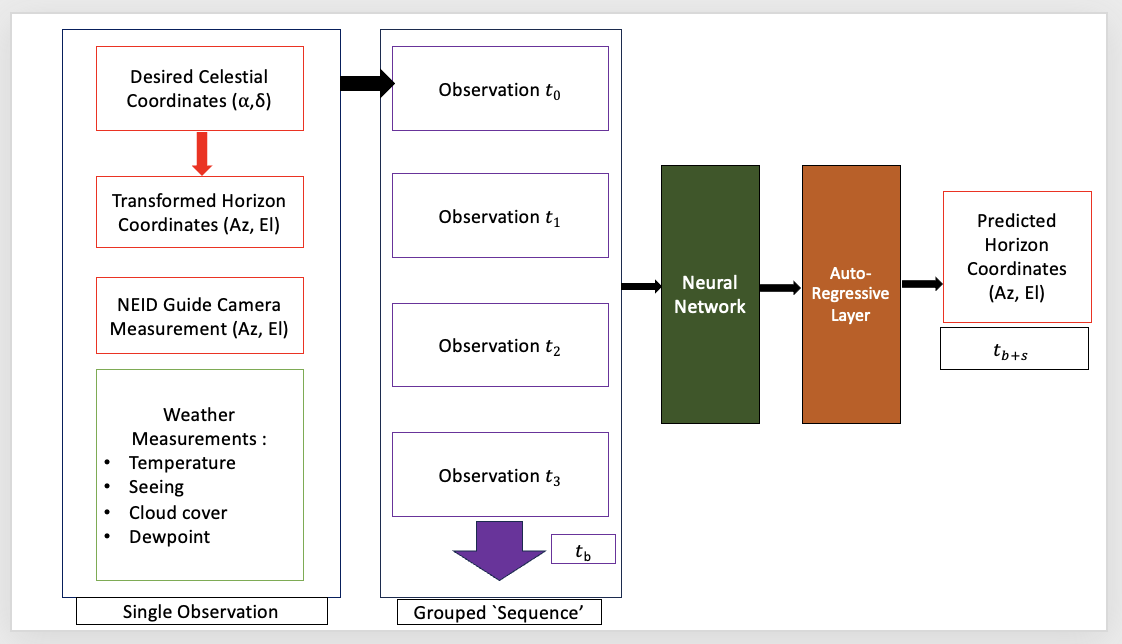}
    \caption{Architecture of the Tracking Network}
    \label{fig:tracknetoutline}
\end{figure}

In order to mimic how the tracking process would function on sky, we split up our training and testing data into input and output sequences separated by a time lag. If one piece of a synthetic path $P$ can be represented as the ordered set $P_a^b=\{\textbf{p}_i\}_{i=a}^{b}$ where each $\textbf{p}_i$ is an observation array of $k$ features at time $i$, we separate our data in a sliding window with 
\begin{equation}
\label{eq:sets}
I=\bigcup_{i=0}^{n-b}P_i^{i+b}, ~~~~~ O=\bigcup_{i=s+b}^n(Az, El)_i,
\end{equation}
where $I$ and $O$ are the input and output data sets respectively. Here, $b$ is the length of the sequences of observations we use as inputs into our model, while $s$ is the delay between the end of the input sequence and the  time-step for which we are trying to predict the optimal horizon coordinates. With this system, we are able to use outputs and corrected measurements from previous observations as features to predict future time-steps. While the NEID system currently has latency of order $s< 5$, to be better able to argue for generalization on other systems we conservatively set $s=b=5$. This delay accounts for the combined latency of measuring offsets from the NEID guide camera, model processing time, as well as the deployment of an additional corrective layer that incorporates recent model error to correct the pre-trained neural network. This additional layer is auto-regressive--a term used to refer to a deep learning framework where previous predictions are used to inform future outputs. It's important to note that this is different from the recurrent architectures described in Sec. \ref{sec:dlback}, which refers only to inputs being sequences of observations. Often auto-regressive techniques are used in conjunction with recurrent networks, as we do here. Fig.\ref{fig:tracknetoutline} shows the total outline of this network in more detail.

\begin{figure}
    \centering
    \includegraphics[scale=0.5]{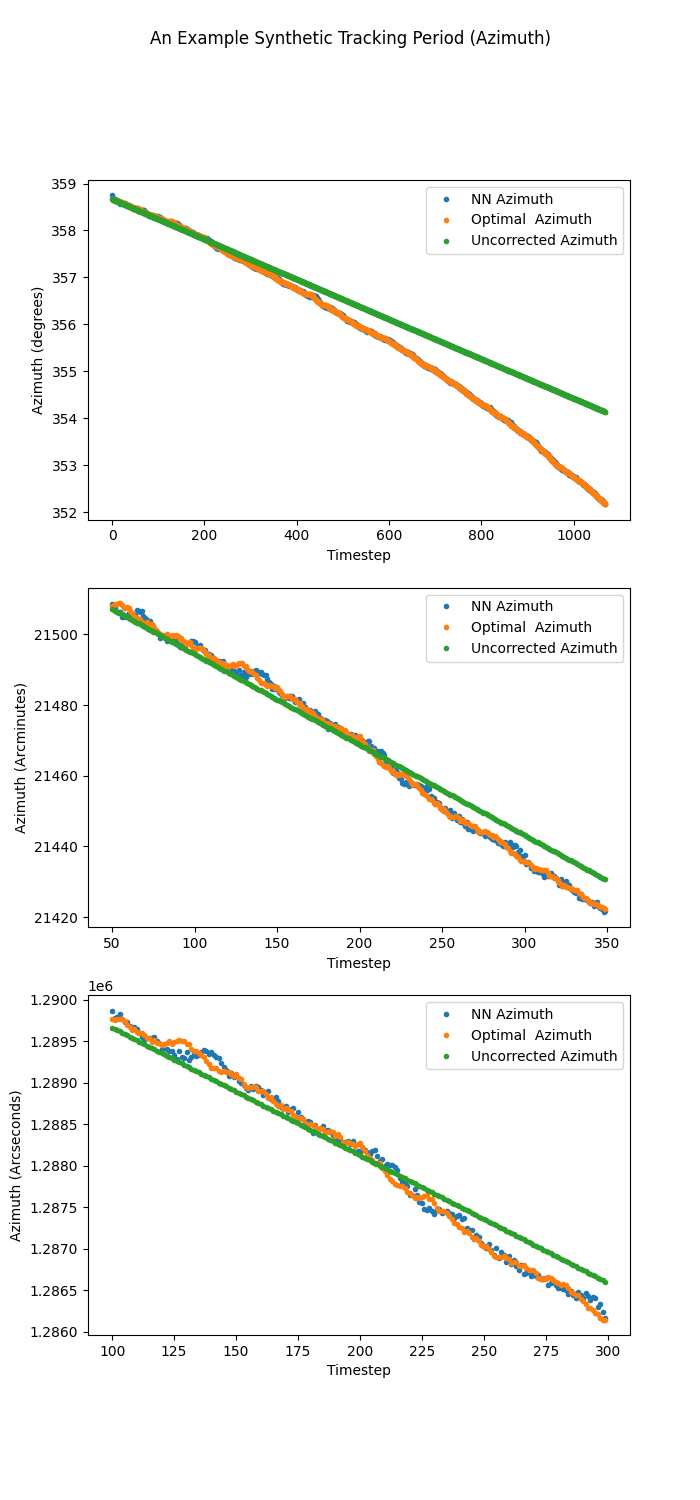}
    \caption{A visualization of our model for predictive tracking using one path sample in our testing set. These examples indicate the success of our model at forecasting complex patterns. Again, the uncorrected azimuth (green) comes directly from the coordinate transform done with astropy.  The ''true" azimuth is that of the target with the addition systematic error introduced by our synthetic features (orange), and the NN azimuth (blue) is our network's forecasted prediction.}
    \label{fig:longgraph}
\end{figure}

In order to gauge the effectiveness of these models, we look at the value representing the 95th percentile of the percentage error in offsets in azimuth and elevation. This figure tracks overall performance, minimally biased by outliers. Table \ref{tab:rectable} displays these percentile values for the recurrent networks on the testing set. In addition, we compare with the astropy coordinate transforms, which serve as our telescope independent benchmark. Finally we show the performance of a shallow linear model. For a detailed look at one example tracking path, Fig. \ref{fig:longgraph} displays a subset of a single ``observation" or tracking path, illustrating the performance of our predictive GRU network. We show the performance on three scales: degrees, arcminutes, and arcseconds. The intentionally introduced `systematic' errors and those relating to atmospheric conditions in our synthetic data results in shifts from the baseline astropy coordinate transforms at the level of several arcminutes (as indicated in the first row of Table \ref{tab:rectable}). The synthetic data in our model includes some random noise on the order of a few arcseconds (depending on the length of the path), so we expect our accuracy to be of this same order.
\begin{table}[ht]
\caption{Recurrent Network Performance Statistics} 
\label{tab:rectable}
\begin{center}       
\begin{tabular}{|l|l|l|l|l|} 
\hline
\rule[-1ex]{0pt}{3.5ex} Model & Azimuth $P_{95}$ Percent Error & Elevation $P_{95}$ Percent Error & Az ($''$) & El ($''$) \\
\hline
\rule[-1ex]{0pt}{3.5ex}  Astro Baseline & 6.5075 & 6.4752 & 234.27 & 233.10  \\
\hline
\rule[-1ex]{0pt}{3.5ex}  Linear Baseline & 0.1655 & 0.1312 & 5.96 & 4.72\\
\hline
\rule[-1ex]{0pt}{3.5ex}  LSTM & 0.0850 & 0.0887  & 3.06 & 3.19\\
\hline
\rule[-1ex]{0pt}{3.5ex}  GRU & 0.0610 & 0.0702 & 2.20 & 2.53\\
\hline 
\end{tabular}
\end{center}
\end{table}

\subsection{Tracking Corrections with Other Architectures}
In addition to the recurrent neural networks described above, we also conduct some preliminary testing of other network architectures. These include the temporal convolutional network (TCN), an extrapolation of the common convolutional network used often in image processing, and the transformer network, whose self-attention mechanisms have gained increasing popularity in recent years \cite{nn-book}. In Table \ref{tab:otable} we see the results of these other architectures on the same testing set used in the recurrent case.

\begin{table}[ht]
\caption{Other Network Performance Statistics} 
\label{tab:otable}
\begin{center}       
\begin{tabular}{|l|l|l|l|l|} 
\hline
\rule[-1ex]{0pt}{3.5ex} Model & Azimuth $P_{95}$ Percent Error & Elevation $P_{95}$ Percent Error & Az ($''$) & El ($''$) \\
\hline
\rule[-1ex]{0pt}{3.5ex}  Astro Baseline & 6.5075 & 6.4752 & 234.27 & 233.10  \\
\hline
\rule[-1ex]{0pt}{3.5ex}  Linear Baseline & 0.1655 & 0.1312 & 5.96 & 4.72\\
\hline
\rule[-1ex]{0pt}{3.5ex}  TCN & 0.2924 & 0.1260  & 10.53 & 4.54\\
\hline
\rule[-1ex]{0pt}{3.5ex}  Transformer & 0.3098 & 0.2747 & 11.15 & 9.89\\
\hline 
\rule[-1ex]{0pt}{3.5ex}  GRU-TCN & 0.0817& 0.0919 & 2.94 & 3.31\\
\hline 
\end{tabular}
\end{center}
\end{table}
\section{Summary and Future Work}
\label{sec:summary}
We have developed a target acquisition system, built upon a recurrent neural network, that improves upon the WIYN telescope's current pointing solution, when trained and tested on historical data recorded from the Star Tracking camera. Our system can be trained and re-calibrated during non-observational hours, using only science observations, boosting both the time and cost efficiency needed to obtain accurate pointing. We have also developed a model for predictive target tracking, using a similar deep learning architecture. We evaluated its performance on synthetically-generated data, mimicking the errors, trends and noise expected in real NEID data. Overall, 
 our system is able to improve upon the WIYN's current pointing framework through the use of novel deep learning techniques. We are in the process of testing the predictive tracking system using real tracking logs from ongoing NEID observations. In addition, in the upcoming year we hope to fully deploy and test our system alongside TPoint on sky, as we work towards a more generalized model that may be applied to other instruments and telescopes.
\appendix    

\newpage
\bibliographystyle{plain}
\bibliography{report}
\end{document}